\vsize=9.0in\voffset=1cm
\looseness=2


\message{fonts,}

\font\tenrm=cmr10
\font\ninerm=cmr9
\font\eightrm=cmr8
\font\teni=cmmi10
\font\ninei=cmmi9
\font\eighti=cmmi8
\font\ninesy=cmsy9
\font\tensy=cmsy10
\font\eightsy=cmsy8
\font\tenbf=cmbx10
\font\ninebf=cmbx9
\font\tentt=cmtt10
\font\ninett=cmtt9

\font\ninesl=cmsl9
\font\eightsl=cmsl8

\font\nineit=cmti9
\font\eightit=cmti8

\skewchar\ninei='177 \skewchar\eighti='177
\skewchar\ninesy='60 \skewchar\eightsy='60

\def\eightpoint{\def\rm{\fam0\eightrm} 
\normalbaselineskip=9pt
\normallineskiplimit=-1pt
\normallineskip=0pt

\textfont0=\eightrm \scriptfont0=\sevenrm \scriptscriptfont0=\fiverm
\textfont1=\ninei \scriptfont1=\seveni \scriptscriptfont1=\fivei
\textfont2=\ninesy \scriptfont2=\sevensy \scriptscriptfont2=\fivesy
\textfont3=\tenex \scriptfont3=\tenex \scriptscriptfont3=\tenex
\textfont\itfam=\eightit  \def\it{\fam\itfam\eightit} 
\textfont\slfam=\eightsl \def\sl{\fam\slfam\eightsl} 

\setbox\strutbox=\hbox{\vrule height6pt depth2pt width0pt}%
\normalbaselines \rm}

\def\ninepoint{\def\rm{\fam0\ninerm} 
\textfont0=\ninerm \scriptfont0=\sevenrm \scriptscriptfont0=\fiverm
\textfont1=\ninei \scriptfont1=\seveni \scriptscriptfont1=\fivei
\textfont2=\ninesy \scriptfont2=\sevensy \scriptscriptfont2=\fivesy
\textfont3=\tenex \scriptfont3=\tenex \scriptscriptfont3=\tenex
\textfont\itfam=\nineit  \def\it{\fam\itfam\nineit} 
\textfont\slfam=\ninesl \def\sl{\fam\slfam\ninesl} 
\textfont\bffam=\ninebf \scriptfont\bffam=\sevenbf
\scriptscriptfont\bffam=\fivebf \def\bf{\fam\bffam\ninebf} 
\textfont\ttfam=\ninett \def\tt{\fam\ttfam\ninett} 

\normalbaselineskip=11pt
\setbox\strutbox=\hbox{\vrule height8pt depth3pt width0pt}%
\let \smc=\sevenrm \let\big=\ninebig \normalbaselines
\parindent=1em
\rm}

\def\tenpoint{\def\rm{\fam0\tenrm} 
\textfont0=\tenrm \scriptfont0=\ninerm \scriptscriptfont0=\fiverm
\textfont1=\teni \scriptfont1=\seveni \scriptscriptfont1=\fivei
\textfont2=\tensy \scriptfont2=\sevensy \scriptscriptfont2=\fivesy
\textfont3=\tenex \scriptfont3=\tenex \scriptscriptfont3=\tenex
\textfont\itfam=\nineit  \def\it{\fam\itfam\nineit} 
\textfont\slfam=\ninesl \def\sl{\fam\slfam\ninesl} 
\textfont\bffam=\ninebf \scriptfont\bffam=\sevenbf
\scriptscriptfont\bffam=\fivebf \def\bf{\fam\bffam\tenbf} 
\textfont\ttfam=\tentt \def\tt{\fam\ttfam\tentt} 

\normalbaselineskip=11pt
\setbox\strutbox=\hbox{\vrule height8pt depth3pt width0pt}%
\let \smc=\sevenrm \let\big=\ninebig \normalbaselines
\parindent=1em
\rm}

\message{fin format jgr}

\magnification=1200
\font\Bbb=msbm10
\def\R{\hbox{\Bbb R}}
\def\C{\hbox{\Bbb C}}
\def\S{\hbox{\Bbb S}}
\def\N{\hbox{\Bbb N}}
\def\pa{\partial}
\def\ep{\varepsilon}
\def\b{\backslash}

\vskip 4 mm
\centerline{\bf On iterative reconstruction in the nonlinearized polarization
tomography}
\vskip 2 mm
\centerline{\bf R. G. Novikov}
\vskip 2 mm
\centerline{16 April 2009}
\vskip 2 mm

\noindent
{\ninerm CNRS (UMR 7641), Centre de Math\'ematique Appliqu\'ees, Ecole
Polytechnique,}

\noindent
{\ninerm 91128 Palaiseau, France}

\noindent
{\ninerm e-mail: novikov@cmapx.polytechnique.fr}

\vskip 4 mm
{\bf Abstract.} We give uniqueness theorem and reconstruction algorithm for
the non-linearized problem of finding the dielectric anisotropy $f$ of the
medium from non-overdeter-

\noindent
mined polarization tomography data.
We assume that the medium has uniform background parameters and that the
anisotropic (dielectric permeability) perturbation is described
by symmetric and sufficiently small matrix-function $f$. On a pure
mathematical level this article contributes to the theory of
non-abelian Radon transforms and to iterative methods of inverse scattering.

\vskip 4 mm
{\bf 1. Introduction}

We consider the system
$$\theta\pa_x\eta=\pi_{\theta}f(x)\eta,\ \ x\in\R^3,\ \ \theta\in\S^2,
\eqno(1.1)$$
where
$$\eqalignno{
&\theta\pa_x=\sum_{j=1}^3\theta_j{\pa\over \pa x_j},&(1.2)\cr
&\eta\ \ {\rm at\ fixed}\ \  \theta\ \  {\rm is\ a\ function\ on}\ \  \R^3\ \
{\rm with\ values\ in}\ \  Z_{\theta},&(1.3)\cr
&Z_{\theta}=\{z\in\C^3:\ z\theta=0\},&(1.4)\cr}$$
$$\eqalign{
&f\ \ {\rm is\ a\ sufficiently\ regular\ function\ on}\ \ \R^3\ \
{\rm with\ values\ in}\ \  {\cal M}_{3,3}\cr
&{(\rm that\ is\ in}\ \ 3\times 3\ \
{\rm complex\ matrices)\ with\ sufficient\ decay\ at\ infinity},\cr}\eqno(1.5)
$$
$$\pi_{\theta}\ \  {\rm is\ the\ orthogonal\ projector\ on}\ \  Z_{\theta}.
\eqno(1.6)$$
In (1.1) the unit vector $\theta$ is considered as a spectral parameter.

System (1.1) arises in the electromagnetic polarization tomography and is a
system of differential equations for the polarization vector-function $\eta$
in a medium with zero conductivity, unit magnetic permeability and
appropriately small anisotropic perturbation of some uniform dielectric
permeability.
This anisotropic perturbation of the dielectric permeability tensor
corresponds to the matrix-function $f$ of (1.1). In addition, by some
physical arguments, $f$ must be skew-Hermition, $f_{ij}=-\bar f_{ji}$.
For more information on physics of the electromagnetic polarization
tomography see [Sh1], [NS], [Sh3] and references therein (and, in particular,
[KO] and [A]).

Let
$$\eqalignno{
&\omega\in\S^2,\ \ \theta\in\S^1_{\omega},\ \
\theta^{\perp}=\omega\times\theta,&(1.7)\cr
&\mu_1=\eta\omega,\ \ \mu_2=\eta\theta^{\perp}\ \ {\rm for}\ \
\eta\in Z_{\theta},&(1.8)\cr}$$
where
$$\S_{\omega}^1=\{\theta\in\S^2:\ \theta\omega=0\},\eqno(1.9)$$
$\times$ denotes vector product, $Z_{\theta}$ is defined by (1.4).

From (1.1)-(1.9) it follows that
$$\eqalign{
&\theta\pa_x\mu=F(x,\theta,\omega)\mu,\ \ x\in\R^3,\ \ \theta\in\S^1_{\omega},
\cr
&F(x,\theta,\omega)=\pmatrix{
\omega f(x)\omega &\ \omega f(x)\theta^{\perp}\hfill\cr
\theta^{\perp}f(x)\omega &\ \theta^{\perp}f(x)\theta^{\perp}\hfill\cr},\ \
\xi f(x)\zeta=\sum_{1\le i,j\le 3}f_{ij}(x)\xi_i\zeta_j,\cr}
\eqno(1.10)$$
where $\mu$ is related with $\eta$ of (1.1) by (1.8) and is a $\C^2$-valued
function on $\R^3$ for fixed $\omega$ and $\theta$.

We consider also (1.10) for $\mu$ taking its values in ${\cal M}_{2,2}$
(that is in $2\times 2$ complex matrices).
Let $\mu^+$ denote the solution of (1.10)
 such that
$$\lim\limits_{s\to -\infty}\mu^+(x+s\theta,\theta,\omega)=Id\ \ {\rm for}\ \
x\in\R^3,\eqno(1.11)$$
where $Id$ is the $2\times 2$ identity matrix. Let
$$S(x,\theta,\omega)=\lim\limits_{s\to +\infty}\mu^+(x+s\theta,\theta,\omega),
\ \ (x,\theta)\in\Gamma_{\omega},\eqno(1.12)$$
where
$$\eqalign{
&\Gamma_{\omega}=\{(x,\theta):\ x\in X_{\theta},\
\theta\in\S^1_{\omega}\},\ \ \omega\in\S^2,\cr
&X_{\theta}=Re\,Z_{\theta},\ \ \theta\in\S^2,\cr}\eqno(1.13)$$
where $\S^1_{\omega}$ is defined by (1.9), $Z_{\theta}$ is defined by (1.4).
In addition, $\mu^+$ and $S$ are well defined due to (1.5).

Note that
$$\Gamma_{\omega}\subset T\S^2,\ \ \omega\in\S^2,\eqno(1.14)$$
where
$$T\S^{d-1}=\{(x,\theta):\ x\in\R^d,\ \ \theta\in\S^{d-1},\ \ x\theta=0\}.
\eqno(1.15)$$
In addition, we interpret $T\S^{d-1}$ as the set of all rays in $\R^d$. As a
ray $\gamma$ we understand a straight line with fixed orientation. If
$\gamma=(x,\theta)\in T\S^{d-1}$, then

\noindent
$\gamma=\{y\in\R^d:\ y=x+s\theta,\ \
s\in\R\}$ (modulo orientation) and $\theta$ gives the orientation of $\gamma$.
Note also that
$$\Gamma_{\omega}\approx\R^2\times\S^1,\eqno(1.16a)$$
or, more precisely,
$$(x,\theta)\in\Gamma_{\omega}\Leftrightarrow x=\xi_1\theta^{\perp}+
\xi_2\omega,\ \ \xi=(\xi_1,\xi_2)\in\R^2,\ \ \theta\in\S^1_{\omega}\approx
\S^1,\eqno(1.16b)$$
where $\omega\in\S^2$, $\theta^{\perp}=\omega\times\theta$. In addition, we
consider $(\xi,\theta)\in\R^2\times\S^1$ as coordinates on $\Gamma_{\omega}$.

One can see that $S$ of (1.12) is a matrix-function on
$$\eqalign{
&\Sigma=\{(\gamma,\omega):\ \gamma\in\Gamma_{\omega},\ \ \omega\in\S^2\}=\cr
&\{(\gamma,\omega):\ \gamma=(x,\theta)\in T\S^2,\ \ \omega\in\S^1_{\theta}\},
\cr}\eqno(1.17)$$
where $\S^1_{\theta}$ is defined as in (1.9). On the other hand, one can show
that
$$\eqalign{
&S(x,\theta,\omega)\ \ {\rm at\ fixed}\ \ \omega\in\S^1_{\theta}\ \ {\rm and}
\ \ (x,\theta)\in T\S^2\cr
&{\rm uniquely\ determines}\ \ S(x,\theta,\cdot)\ \ {\rm on}\ \ \S^1_{\theta}
\ \ {\rm and},\cr
&{\rm as\ a\ corollary},\ \ S\ \ {\rm can\ be\ considered\ as\ a\
matrix-function\ on}\ \ T\S^2.\cr}\eqno(1.18)$$

The matrix-function $S$ can be considered as a non-abelian ray transform
of $f$. See [MZ], [V], [Sh2], [N], [FU], [E], [M], [DP], [P]
and references therein for some other non-abelian ray
transforms.

In the present work we say that $S$ of (1.11)-(1.13) is the polarization
ray transform of $f$.

Using the terminology of the scattering theory one can say also that $S$ is
the "scattering" matrix for system (1.10).

The basic problem of the polarization tomography in the framework of the
model described by (1.1), (1.10) consists in finding $f$ on $\R^3$ from
$S$ on $\Lambda$, where $\Lambda$ is some appropriate subset of $\Sigma$ of
(1.17). It is especially natural to consider this problem for the case
when $dim\,\Lambda=3$.

From results of [NS] it follows that there is a non-uniqueness in this
problem if $f$ is not symmetric even if $S$  is given on $\Lambda=\Sigma$.
Results of [NS] also imply a local uniqueness theorem (up to a natural
obstruction if $f$ is not symmetric) for the case when $S$ is given on
$\Sigma$ (or on $T\S^2$ in the sense (1.18)).

In the present work we consider the following inverse problem for equations
(1.1), (1.10).

\vskip 2 mm
{\bf Problem 1.1.}
Find symmetric $f$, $f_{ij}=f_{ji}$, from $S$ on $\Lambda$ (or from partial
information about $S$ on $\Lambda$), where
$$\Lambda=\{(\gamma,\omega):\ \gamma\in\Gamma_{\omega},\ \ \omega\in
\{\omega^1,\ldots,\omega^k\}\},\eqno(1.19)$$
where $\omega^1,\ldots,\omega^k$ are some fixed points of $\S^2$.

One can see that Problem 1.1 is a version of the aforementioned basic problem
of the polarization tomography with $dim\,\Lambda=3$, see definitions
(1.13), (1.17), (1.19).

The main results of the present work consist in  uniqueness theorem and
reconstruction algorithm
for nonlinearized Problem 1.1 with sufficiently small $f$, where only
the element $S_{11}$ of $S=(S_{ij})$ on $\Lambda$ is used as the data and
where
$$\eqalign{
&k=6,\ \omega^1=e_1,\ \omega^2=e_2,\ \omega^3=e_3,\cr
&\omega^4=(e_1+e_2)/\sqrt{2},\ \omega^5=(e_1+e_3)/\sqrt{2},\
\omega^6=(e_2+e_3)/\sqrt{2},\cr}\eqno(1.20)$$
where $e_1,e_2,e_3$ is the basis in $\R^3$.  See Sections 2, 3 and,
in particular, Theorem 3.1.

One can see that this reconstruction is non-overdetermined:\ we reconstruct
6 functions $f_{ij}$, $1\le i\le j\le 3$, on $\R^3$ from 6 functions
$S_{11}(\cdot,\omega)$, $\omega\in\{\omega^1,\ldots,\omega^6\}$, of 3
variables.

Our reconstruction is iterative and its first apprximation more or less
coincides with the linearized polarization tomography reconstruction of
Section 5.1 of [Sh1]. In addition, we give estimates on the reconstruction
error $f-f^n$ for the approximation $f^n$ with number $n\in\N$, see
Theorem 3.1 .
To our knowledge even $f-f^1$ was not estimated rigorously in the literature.

The main results of the present work are presented in detail in Sections 2
and 3.

Some possible development of the present work and some open questions are
mentioned in Section 6.

\vskip 2 mm
{\bf 2. Reconstruction algorithm}

Consider the classical ray transform {\it I} defined by the formula
$$If(\gamma)=\int\limits_{\R}f(x+s\theta)ds,\ \
\gamma=(x,\theta)\in T\S^{d-1},\eqno(2.1)$$
for any complex-valued sufficiently regular function $f$ on $\R^d$ with
sufficient decay at infinity, where $T\S^{d-1}$ is defined by (1.15) (and
where $d=2$ or $d=3$).

We use the following Radon-type inversion formula for {\it I} in dimension
$d=2$:
$$\eqalign{
&f(x)={1\over 4\pi}\int\limits_{\S^1}
h^{\prime}(x\theta^{\perp},\theta)d\theta,\ \
h^{\prime}(s,\theta)={d\over ds}h(s,\theta),\cr
&h(s,\theta)={1\over \pi}p.v.\int\limits_{\R}{g(t,\theta)\over {s-t}}dt,\cr}
\eqno(2.2)$$
where $g(s,\theta)=If(s\theta^{\perp},\theta)$,
$x=(x_1,x_2)\in\R^2$, $\theta=(\theta_1,\theta_2)\in\S^1$,
$\theta^{\perp}=(-\theta_2,\theta_1)$, $s\in\R$, $d\theta$ is the standard
element of arc length on $\S^1$.

We use the following slice by slice reconstruction of $f$ on $\R^3$ from
$g=If$ on $\Gamma_{\omega}$ of (1.13) for fixed $\omega\in\S^2$:
$$g\bigg|_{T\S^1(Y)}\buildrel (2.2) \over \to\ \ f\bigg|_Y \eqno(2.3a)$$
for each two-dimensional plane $Y$ of the form
$$Y=X(\S^1_{\omega})+y,\ \ y\in X^{\perp}(\S^1_{\omega}),\eqno(2.3b)$$
where $\S^1_{\omega}$ is defined by (1.9), $X(\S^1_{\omega})$ is the linear
span of\  $\S^1_{\omega}$ in $\R^3$, $X^{\perp}(\S^1_{\omega})$ is the
orthogonal complement of $X(\S^1_{\omega})$ in $\R^3$, $T\S^1(Y)$ is the
set of all oriented straight lines lying in $Y$. In addition,
$$\Gamma_{\omega}=
\cup_{y\in X^{\perp}(\S^1_{\omega})}T\S^1(X(\S^1_{\omega})+y).
\eqno(2.4)$$

Consider the three-dimensional transverse ray transformation ${\it J}$
 defined by the formula (see Section 5.1 of [Sh1]):
$$\eqalign{
&Jf(\gamma,\omega)=I(\omega f\omega)(\gamma)=\cr
&\int\limits_{\R}\omega f(x+s\theta)\omega ds,\ \ (\gamma,\omega)\in\Sigma,\ \
 \gamma=(x,\theta),\cr}\eqno(2.5)$$
for any ${\cal M}_{3,3}$-valued sufficiently regular function $f$ on $\R^3$
with sufficient decay at infinity, where $\omega f\omega$ is defined as in
(1.10), $\Sigma$ is defined by (1.17).

We use the following reconstruction of symmetric $f$ on $\R^3$ from $Jf$ on
$\Lambda$ of (1.19) for $\omega^1,\ldots,\omega^k$ given by (1.20):
$$\eqalign{
&f_{jj}=I^{-1}_{\omega^j}g_{\omega^j},\ \ j=1,2,3,\cr
&f_{12}=I^{-1}_{\omega^4}g_{\omega^4}-{1\over 2}(f_{11}+f_{22}),\cr
&f_{13}=I^{-1}_{\omega^5}g_{\omega^5}-{1\over 2}(f_{11}+f_{33}),\cr
&f_{23}=I^{-1}_{\omega^6}g_{\omega^6}-{1\over 2}(f_{22}+f_{33}),\cr}\eqno(2.6)
$$
where $g_{\omega}=Jf\big|_{\Gamma_{\omega}}$ and
$I^{-1}_{\omega}$ denotes
the slice by slice reconstruction via inversion formulas (2.2), (2.3) for
${\it I}$ from data on $\Gamma_{\omega}$.

Now we are ready to present our iterative reconstruction of sufficiently
small, symmetric and compactly supported $f$ from the element $S_{11}$ of
$S=(S_{ij})$ on $\Lambda$, where $S$ is defined by (1.12), $\Lambda$ is
defined by (1.19), (1.20).

Thus, in addition to (1.5), we assume that
$$\eqalign{
&f\ \ {\rm is\ symmetric},\ \ f_{ij}=f_{ji},\cr
&f(x)\equiv 0\ \ {\rm for}\ \ |x|\ge r_0,\cr
&f\ \ {\rm is\ sufficiently\ small}.\cr}\eqno(2.7)$$

Let
$$\eqalignno{
&\Delta^0=(S_{11}-1)\big|_{\Lambda},&(2.8)\cr
&f^1=\chi J^{-1}_{\Lambda}\Delta^0,&(2.9)\cr}$$
where $J^{-1}_{\Lambda}$ denotes the reconstruction via inversion formulas
(2.6) for $J$ from data on $\Lambda$, $\chi$ denotes the multiplication by
smooth $\chi$ such that
$$\eqalign{
&\chi(x)\equiv 1\ \ {\rm for}\ \ |x|\le r_0,\cr
&\chi(x)\equiv 0\ \ {\rm for}\ \ |x|\ge r_1,\cr}\eqno(2.10)$$
where $r_0$ is the number of (2.7), $r_1>r_0$.

In our iterative reconstruction, $f^1$ is the first approximation to $f$.

From the approximation $f^n$ with number $n$ the approximation $f^{n+1}$ with
number $n+1$ is constructed as follows:

(1) We find the element $\mu_{11}^{n+}$ of $\mu^{n+}=(\mu_{ij}^{n+})$ on
$${\cal V}=\{(x,\theta,\omega):\ x\in\R^3,\ \theta\in\S^1_{\omega},\
\omega\in\{\omega^1,\ldots,\omega^6\}\},\eqno(2.11)$$
where $\mu^{n+}$ satisfies (1.10), (1.11) with $f^n$ in place of $f$ in
(1.10);

(2) We find
$$\eqalignno{
&S^n_{11}(x,\theta,\omega)=\lim\limits_{s\to +\infty}
\mu_{11}^{n+}(x+s\theta,\theta,\omega),\ \ (x,\theta,\omega)\in\Lambda,
&(2.12)\cr
&\Delta^n=(S_{11}-S^n_{11})\big|_{\Lambda};&(2.13)\cr}$$

(3) Finally, we find
$$f^{n+1}=\chi(f^n+J^{-1}_{\Lambda}\Delta^n),\eqno(2.14)$$
where $J^{-1}_{\Lambda}$ and $\chi$ are the same that in (2.9).

Note that in (2.9), (2.14) we do not assume that $\Delta^0$, $\Delta^n$ are
in the range of $J$. However, $J_{\Lambda}^{-1}g$ is well-defined on the
basis of (2.6) for any
$$\eqalign{
&g=(g_{\omega^1},\ldots,g_{\omega^6}),\ \ {\rm where}\cr
&g_{\omega^i}\ \ {\rm is\ a\ complex-valued\ sufficiently\ regular}\cr
&{\rm function\ on}\ \ \Gamma_{\omega^i}\ \ {\rm with\ sufficient\ decay\ at}
\cr
&{\rm infinity\ (see\ (1.16))\ for\ each}\ \ i\in\{1,\ldots,6\}.\cr}
\eqno(2.15)$$

\vskip 2 mm
{\bf 3. Convergence}

We consider
$$\hat L^{\infty,\sigma}(\R^3)=\{u:\ \hat u\in L^{\infty}(\R^3),\ \
\|u\|_{\hat L^{\infty,\sigma}(\R^3)}<+\infty\},\ \ \sigma\ge 0,\eqno(3.1)$$
where
$$\hat u(p)=\bigl({1\over 2\pi}\bigr)^3\int\limits_{\R^3}e^{ipx}u(x) dx,\ \
p\in\R^3,\eqno(3.2)$$
$$\eqalign{
&\|u\|_{\hat L^{\infty,\sigma}(\R^3)}=\|\hat u\|_{L^{\infty,\sigma}(\R^3)},
\cr
&\|\hat u\|_{L^{\infty,\sigma}(\R^3)}=ess\,\sup_{p\in\R^3}
(1+|p|)^{\sigma}|\hat u(p)|.\cr}\eqno(3.3)$$
We consider
$$\hat C^{\alpha,\sigma}(\R^3)=\{u:\ \hat u\in C^{[\alpha]}(\R^3),\ \
\|u\|_{\hat C^{\alpha,\sigma}(\R^3)}< +\infty\},\ \alpha\ge 0,\
\sigma\ge 0,\eqno(3.4)$$
where $\hat u$ is defined by (3.2), $C^{[\alpha]}$ denotes $[\alpha]$-times
continuously differentiable functions, $[\alpha]$ is the integer part of
$\alpha$,
$$\eqalignno{
&\|u\|_{\hat C^{\alpha,\sigma}(\R^3)}=
\|\hat u\|_{C^{\alpha,\sigma}(\R^3)},&(3.5)\cr
&\|\hat u\|_{C^{\alpha,\sigma}(\R^3)}=\sup_{|J|\le [\alpha],\ p\in\R^3}
(1+|p|)^{\sigma}|\pa^J\hat u(p)|\ \ {\rm for}\ \ \alpha=[\alpha],&(3.6)\cr}$$
$$\eqalign{
&\|\hat u\|_{C^{\alpha,\sigma}(\R^3)}=\max\,(N_1,N_2)\ \ {\rm for}\ \
\alpha> [\alpha],\cr
&N_1=\|\hat u\|_{C^{[\alpha],\sigma}(\R^3)},\cr
&N_2=\sup_{|J|=[\alpha],\ p\in\R^3,\ p^{\prime}\in\R^3,\ |p-p^{\prime}|\le 1}
(1+|p|)^{\sigma}{|\pa^J\hat u(p^{\prime})-\pa^J\hat u(p)|\over
|p-p^{\prime}|^{\alpha-[\alpha]}},\cr}\eqno(3.7)$$
where
$$\pa^J\hat u(p)={\pa^{|J|}\hat u(p)\over
\pa p_1^{J_1}\pa p_2^{J_2}\pa p_3^{J_3}},\ \  J=(J_1,J_2,J_3)\in (\N\cup 0)^3,
\ |J|=J_1+J_2+J_3.\eqno(3.8)$$

In addition, in (3.1)-(3.8) we assume that $u$, $\hat u$ are
${\cal M}_{n_1,n_2}$-valued functions, in general, where
${\cal M}_{n_1,n_2}$ is the space of $n_1\times n_2$ matrices with
complex elements,
$$|M|=\max_{\scriptstyle 1\le i\le n_1\atop 1\le j\le n_2}|M_{ij}|\ \
{\rm for}
\ \ M\in {\cal M}_{n_1,n_2}.\eqno(3.9)$$
In addition, in the present work we always have that
$1\le n_1\le 3$, $1\le n_2\le 3$.

\vskip 2 mm
{\bf Lemma 3.1.} {\it Let} $u\in\hat L^{\infty,\sigma}(\R^3)$,
$v\in\hat C^{\alpha,\sigma}(\R^3)$, {\it where} $\alpha\ge 0$, $\sigma>3$.
{\it In addition, in general, we assume that} $u$ {\it is}
${\cal M}_{n_1,n_2}$- {\it valued and} $v$ {\it is}
${\cal M}_{m_1,m_2}$- {\it valued, where} $m_2=n_1$ {\it or/and}
$n_2=m_1$ ({\it and where} $1\le n_1,n_2,\ m_1,m_2\le 3$).
{\it Let}
$$\eqalign{
&{\it either}\ w=v u\ \ {\it for}\ \ m_2=n_1\cr
&{\it or}\ w=u v\ \ {\it for}\ \ n_2=m_1.\cr}\eqno(3.10)$$
{\it Then for each of} $w$ {\it of} (3.10) {\it the following estimate holds:}
$$\eqalign{
&w\in\hat C^{\alpha,\sigma}(\R^3)\cr
&\|w\|_{\hat C^{\alpha,\sigma}(\R^3)}\le\lambda_1(\alpha,\sigma)
\|v\|_{\hat C^{\alpha,\sigma}(\R^3)}
\|u\|_{\hat L^{\infty,\sigma}(\R^3)}\cr}\eqno(3.11)$$
{\it for some positive} $\lambda_1=\lambda_1(\alpha,\sigma)$.

Lemma 3.1 is proved in Section 4.

We consider
$$\hat L^{\infty,\sigma}(\Lambda)=\{g:\ \hat g\in L^{\infty}(\Lambda),\ \
\|g\|_{\hat L^{\infty,\sigma}(\Lambda)}< +\infty\},\ \ \sigma\ge 0,\eqno(3.12)
$$
where $g$ is complex-valued,
$\Lambda$ is defined by (1.19) with $\omega^1,\ldots,\omega^k$ given by
(1.20),
$$\eqalign{
&g=(g_{\omega^1},\ldots,g_{\omega^k}),\ \
\hat g=(\hat g_{\omega^1},\ldots,\hat g_{\omega^k}),\cr
&g_{\omega^i}=g\big|_{\Gamma_{\omega^i}},\ \
\hat g_{\omega^i}=\hat g\big|_{\Gamma_{\omega^i}},\cr
&\hat g_{\omega^i}(p,\theta)=\bigl({1\over 2\pi}\bigr)^2
\int\limits_{X_{\theta}}e^{ipx}g_{\omega^i}(x,\theta)dx,\ \
(p,\theta)\in\Gamma_{\omega^i},\ \ i=1,\ldots,k,\cr}\eqno(3.13)$$
$$\eqalign{
&\|g\|_{\hat L^{\infty,\sigma}(\Lambda)}=
\|\hat g\|_{L^{\infty,\sigma}(\Lambda)},\cr
&\|\hat g\|_{L^{\infty,\sigma}(\Lambda)}=\max_{i\in\{1,\ldots,k\}}
ess\,\sup_{(p,\theta)\in\Gamma_{\omega^i}}(1+|p|)^{\sigma}
|\hat g_{\omega^i}(p,\theta)|,\cr}\eqno(3.14)$$
where $\Gamma_{\omega}$ and $X_{\theta}$ are defined according to (1.13).
Actually, in (3.12) we consider $L^{\infty}(\Lambda)$ as
$$L^{\infty}(\Lambda)=L^{\infty}(\Gamma_{\omega^1})\oplus\ldots\oplus
L^{\infty}(\Gamma_{\omega^k}).\eqno(3.15)$$

We assume that
$$\eqalign{
&f\ \ {\rm is\ a}\ \ {\cal M}_{3,3}-\ {\rm valued\ function\ on}\ \ \R^3,\cr
&f\in\hat L^{\infty,\sigma}(\R^3)\ \ {\rm for\ some}\ \ \sigma>3,\cr
&f\ \ {\rm is\ symmetric},\ \ f_{ij}=f_{ji},\cr
&f(x)\equiv 0\ \ {\rm for}\ \ |x|\ge r_0,\cr}\eqno(3.16)$$
$$\eqalign{
&\chi\ \  {\rm is\ a\ nonnegative\ real-valued\ function\ on}\ \ \R^3,\cr
&\chi\in C^m(\R^3)\ \ {\rm for\ some}\ \ m\in\N,\ m\ge\sigma,\cr
&\chi(x)\ge\chi(y)\ \ {\rm if}\ |y|\ge |x|,\cr
&\chi(x)\equiv 1\ \ {\rm for}\ \ |x|\le r_0,\cr
&\chi(x)\equiv 0\ \ {\rm for}\ \ |x|\ge r_1,\cr}\eqno(3.17)$$
where $r_0$, $r_1$ are some fixed real numbers, $r_0<r_1$.
Properties (3.16), (3.17) imply, in particular, that
$$\eqalign{
&\chi f=f,\cr
&\chi\in \hat C^{\alpha,\sigma}(\R^3)\ \ {\rm for\ any}\ \ \alpha\ge 0.
\cr}\eqno(3.18)$$

\vskip 2 mm
{\bf Theorem 3.1.}
{\it Let} $f$ {\it and} $\chi$ {\it satisfy} (3.16), (3.17). {\it Let}
$$\eqalign{
&\|f\|_{\hat L^{\infty,\sigma}(\R^3)}\le\ep\le\ep_0(\alpha,\sigma,\rho),\cr
&\|\chi\|_{\hat C^{1+\alpha,\sigma}(\R^3)}\le\rho\cr}\eqno(3.19)$$
{\it for some} $\alpha\in ]0,1[$, $\rho>0$, $\ep_0>0$, {\it where}
$\ep_0=\ep_0(\alpha,\sigma,\rho)$ {\it is sufficiently small. Let} $S$
{\it be
the polarization ray transform of} $f$ ({\it see Section 1 and,
in particular,
 formulas (1.11), (1.12)). Let } $\Lambda$ {\it be defined by} (1.19), (1.20).
{\it Then the element} $S_{11}$ {\it of} $S=(S_{ij})$ {\it on} $\Lambda$
{\it uniquely determines} $f$ {\it by the iterative reconstruction algorithm
of Section 2; in addition,}
$$\eqalign{
&f^n\to f\ \ {\it in}\ \ \hat L^{\infty,\sigma}(\R^3)\ \ {\it as}\ \
n\to +\infty,\cr
&\|f-f^n\|_{\hat L^{\infty,\sigma}(\R^3)}\le a(\alpha,\sigma,\rho)
(b(\alpha,\sigma,\rho))^{n-1}\ep^{n+1},\cr}\eqno(3.20)$$
{\it for some} $a$ {\it and} $b$, {\it where} $f^n$, $n\in\N$, {\it are
defined by} (2.9), (2.14).

Theorem 3.1 follows from Lemma 3.1, properties (3.18), and Propositions 3.1,
3.2, 3.3.

\vskip 2 mm
{\bf Proposition 3.1.}
{\it Let}
$$g\in\hat L^{\infty,\sigma}(\Lambda)\ \ {\it for\ some}\ \ \sigma>3,
\eqno(3.21)$$
{\it where} $g$ {\it is complex-valued (and} $\Lambda$ {\it is defined by}
(2.19), (2.20)). {\it Let} $J^{-1}_{\Lambda}$ {\it be defined as in} (2.9).
{\it Then}
$$\eqalign{
&J^{-1}_{\Lambda}g\in\hat L^{\infty,\sigma}(\R^3),\cr
&\|J^{-1}_{\Lambda}g\|_{\hat L^{\infty,\sigma}(\R^3)}\le\lambda_2
\|g\|_{\hat L^{\infty,\sigma}(\Lambda)}\cr}\eqno(3.22)$$
{\it for some positive} $\lambda_2$. {\it In addition, if}\ \  $g=J f$,
{\it where} $f$
{\it satisfies} (3.16), {\it then} $g$ {\it satisfies} (3.21) {\it and}
$$J^{-1}_{\Lambda}g=f.\eqno(3.23)$$

Proposition 3.1 is proved in Section 4.

\vskip 2 mm
{\bf Proposition 3.2.}
{\it Let the assumptions of Theorem 3.1 be fulfilled. Let} $\delta^0Jf$
{\it on} $\Lambda$ {\it and} $\delta^1f$ {\it on} $\R^3$ {\it be defined by
the formulas}
$$\eqalignno{
&\Delta^0=Jf+\delta^0Jf\ \ {\it on}\ \ \Lambda,&(3.24)\cr
&J^{-1}_{\Lambda}\Delta^0=f+\delta^1f\ \ {\it on}\ \ \R^3,&(3.25)\cr}$$
{\it where} $\Delta^0$, $J^{-1}_{\Lambda}\Delta^0$ {\it are defined as in}
(2.8), (2.9). {\it Then}
$$\eqalign{
&\delta^0Jf\in\hat L^{\infty,\sigma}(\Lambda),\cr
&\|\delta^0Jf\|_{\hat L^{\infty,\sigma}(\Lambda)}\le
\lambda_3(\alpha,\sigma,\rho)\ep^2,\cr}\eqno(3.26)$$
$$\eqalign{
&\delta^1f\in\hat L^{\infty,\sigma}(\R^3),\cr
&\|\delta^1f\|_{\hat L^{\infty,\sigma}(\R^3)}\le\lambda_2
\lambda_3(\alpha,\sigma,\rho)\ep^2,\cr}\eqno(3.27)$$
{\it for some positive} $\lambda_3=\lambda_3(\alpha,\sigma,\rho)$.

Proposition 3.2 is proved in Section 4.

\vskip 2 mm
{\bf Proposition 3.3.}
{\it Let}
$$f^n=\chi(f+\delta^nf),\eqno(3.28)$$
{\it where} $f$, $\chi$ {\it satisfy} (3.16), (3.17),
$$\eqalign{
&\delta^nf\in\hat L^{\infty,\sigma}(\R^3),\cr
&\|f\|_{\hat L^{\infty,\sigma}(\R^3)}\le\ep\le
\ep_1(\alpha,\sigma,\rho),\cr
&\|f+\delta^nf\|_{\hat L^{\infty,\sigma}(\R^3)}\le\ep\le
\ep_1(\alpha,\sigma,\rho),\cr
&\|\chi\|_{\hat C^{1+\alpha,\sigma}(\R^3)}\le\rho\cr}\eqno(3.29)$$
{\it for some} $\alpha\in ]0,1[$, $\rho>0$, $\ep_1>0$, {\it where}
$\ep_1=\ep_1(\alpha,\sigma,\rho)$ {\it is sufficiently small. Let }
$f^{n+1}$ {\it be constructed from} $S_{11}$ ({\it for} $f$) {\it and}
$f^n$ {\it as described in Section 2. Then}
$$f^{n+1}=\chi(f+\delta^{n+1}f),\eqno(3.30)$$
{\it where}
$$\eqalign{
&\delta^{n+1}f\in\hat L^{\infty,\sigma}(\R^3),\cr
&\|\delta^{n+1}f\|_{\hat L^{\infty,\sigma}(\R^3)}\le
\lambda_4(\alpha,\sigma,\rho)\ep
\|\delta^nf\|_{\hat L^{\infty,\sigma}(\R^3)}\cr}\eqno(3.31)$$
{\it for some positive} $\lambda_4=\lambda_4(\alpha,\sigma,\rho)$.

Proposition 3.3 is proved in Section 5.

To obtain Theorem 3.1 we assume that $\ep_0$ of (3.19) and $\ep_1$ of
(3.29) are so small that
$$\eqalign{
&\ep_0+\lambda_2\lambda_3\ep_0^2\le\ep_1,\ \ \lambda_4\ep_1<1,\cr
&\lambda_4(1+\lambda_2\lambda_3\ep_0)\le b,\ \ b\ep_0<1,\cr}\eqno(3.32)$$
for some $b=b(\alpha,\sigma,\rho)$, where
$\lambda_2$, $\lambda_3$, $\lambda_4$, $\ep_1$ are the constants of
Propositions 3.1, 3.2, 3.3. Under these assumptions, Theorem 3.1 follows
directly from Propositions 3.2, 3.3 and Lemma 3.1.
In addition, we use Proposition  3.3 with $\ep$ given as
$\ep+\lambda_2\lambda_3\ep^2$ in terms of $\ep$ of Theorem 3.1.

\vskip 2 mm
{\bf 4. Proofs of Lemma 3.1 and Propositions 3.1 and 3.2}

Let
$$L^{\infty,\sigma}(\R^3)=\{\hat u\in L^{\infty}(\R^3):\ \
\|\hat u\|_{L^{\infty,\sigma}(\R^3)}< +\infty\},\ \ \sigma\ge 0,\eqno(4.1)$$
where
$\|\cdot\|_{L^{\infty,\sigma}(\R^3)}$ is defined as in (3.3),
$$C^{\alpha,\sigma}(\R^3)=\{\hat u\in C^{[\alpha]}(\R^3):\ \
\|\hat u\|_{C^{\alpha,\sigma}(\R^3)}< +\infty\},\ \ \alpha\ge 0,\
\sigma\ge 0,\eqno(4.2)$$
where $C^{[\alpha]}$ and $\|\cdot\|_{C^{\alpha,\sigma}(\R^3)}$ are defined
as in (3.4) and (3.6), (3.7),
$$L^{\infty,\sigma}(\Lambda)=\{\hat g\in L^{\infty}(\Lambda):\ \
\|\hat g\|_{L^{\infty,\sigma}(\Lambda)}< +\infty\},\ \
\sigma\ge 0,\eqno(4.3)$$
where $\|\cdot\|_{L^{\infty,\sigma}(\Lambda)}$ is defined as in (3.14).
In addition, we assume that $\hat u$ of (4.1), (4.2) is matrix-valued
(of some fixed size), in general, and $\hat g$ of (4.3) is complex-valued.

\vskip 2 mm
{\it Proof of Lemma 3.1.}
We use, in particular, that
$$\eqalign{
&\widehat{u_1u_2}=\hat u_1*\hat u_2,\cr
&\hat u_1*\hat u_2(p)=\int\limits_{\R^3}\hat u_1(p-p^{\prime})
\hat u_2(p^{\prime})dp^{\prime},\ \ p\in\R^3,\cr}\eqno(4.4)$$
where $u_1$, $u_2$ are test functions on $\R^3$, $\hat u$ is defined by (3.2).
 In addition, $u_1$, $u_2$ are matrix-valued, in general, where the matrix
product is defined in the standard way.

By the assumptions of Lemma 3.1  we have that
$$\hat u\in L^{\infty,\sigma}(\R^3),\ \hat v\in  C^{\alpha,\sigma}(\R^3).
\eqno(4.5)$$
Formulas (3.11) follow from (4.4), (4.5), where we use, in particular,
that
$$\int\limits_{\R^3}{dp^{\prime}\over
(1+|p-p^{\prime}|)^{\sigma}(1+|p^{\prime}|)^{\sigma}}\le
{c_1(\sigma)\over (1+|p|)^{\sigma}}\ \ {\rm for}\ \ \sigma>3,\eqno(4.6)$$
for some positive $c_1=c_1(\sigma)$.

Lemma 3.1 is proved.

\vskip 2 mm
{\it Proof of  Proposition 3.1.}
We use that $I^{-1}_{\omega}g_{\omega}$ of (2.3), (2.6) can be defined also
as
$$\eqalign{
&I^{-1}_{\omega}g_{\omega}=u_{\omega},\ \ {\rm where}\cr
&(4\pi)^{-1}(\hat g_{\omega}(p,\theta)+\hat g_{\omega}(p,-\theta))=
\hat u_{\omega}(p),\ \ p\in X_{\theta},\ \ \theta\in\S^1_{\omega},\cr}
\eqno(4.7)$$
where  $u_{\omega}$ and $\hat u_{\omega}$ are related by (3.2),
$g_{\omega}$ and $\hat g_{\omega}$ are related as in (3.13).
Indeed, (4.7) means that
$$\eqalign{
&(2\pi)^{-3}\int\limits_{X_{\theta}}e^{ipx}g_{\omega}^{sym}(x,\theta)dx=
(2\pi)^{-3}\int\limits_{\R^3}e^{ipx}u_{\omega}(x)dx,\cr
&p\in X_{\theta}\b \{x=s\omega,s\in\R\},\ \ \theta\in\S_{\omega}^1,\ \
\omega\in\S^2,\cr}\eqno(4.8)$$
where
$$g_{\omega}^{sym}(x,\theta)={1\over 2}(g_{\omega}(x,\theta)+g(x,-\theta)),\ \
 (x,\theta)\in\Gamma_{\omega}.\eqno(4.9)$$
Representing $p$ in (4.8) as
$$p=p_1\theta^{\perp}+p_2\omega \eqno(4.10a)$$
and representing $x$ in the left hand side of (4.8) as
$$x=\xi_1\theta^{\perp}+\xi_2\omega \eqno(4.10b)$$
and integrating (4.8) with  $e^{-ip_2\xi_2^{\prime}}$, one can see that
$u_{\omega}$ of (4.7) is the function such that
$$\eqalign{
&(2\pi)^{-2}\int\limits_{\R^2}
e^{i\xi_1p_1}g_{\omega}^{sym}(\xi_1\theta^{\perp}+\xi_2\omega,\theta)
\delta(\xi_2-\xi_2^{\prime})d\xi_1d\xi_2=\cr
&(2\pi)^{-2}\int\limits_{\R^3}
e^{ip_1\theta^{\perp}x}u_{\omega}(x)\delta(\omega x-\xi_2^{\prime})dx,\ \
p_1\in\R\b 0,\ \ \xi_2^{\prime}\in\R.\cr}\eqno(4.11)$$
It remains to note that it is actually well-known that the determination of
$u_{\omega}\big|_Y$ from  $g_{\omega}\big|_{TS^1(Y)}$ via (4.9), (4.11) is
equivalent to such a determination on the basis of (2.3), where
$y=\xi_2^{\prime}\omega$.

Formulas (4.7) imply that
$$\eqalign{
&\hat u_{\omega}\in L^{\infty,\sigma}(\R^3),\cr
&\|\hat u_{\omega}\|_{L^{\infty,\sigma}(\R^3)}\le (4\pi)^{-1}
\|\hat g_{\omega}\|_{L^{\infty,\sigma}(\Gamma_{\omega})}\cr
&{\rm if}\ \ \hat g_{\omega}\in L^{\infty,\sigma}(\Gamma_{\omega}),\cr}
\eqno(4.12)$$
where $L^{\infty,\sigma}(\Gamma_{\omega})$ is considered as
$L^{\infty,\sigma}(\Lambda)$ (see (4.3)) for the case when $\Lambda$ is
reduced to the single $\Gamma_{\omega}$.

Proposition 3.1 follows from (2.6) and (4.12).

\vskip 2 mm
{\it Proof of Proposition 3.2.}
We consider $I_{\theta}$, $D_{\theta}$ defined by
$$\eqalignno{
&I_{\theta}u(x)=\int\limits_{\R}u(x+s\theta) ds,\ \ x\in X_{\theta},\ \
\theta\in\S^2,&(4.13)\cr
&D_{\theta}u(x)=\int\limits_0^{+\infty}u(x+s\theta) ds,\ \ x\in\R^3,\ \
\theta\in\S^2,&(4.14)\cr}$$
where $u$ is a matrix-valued test function on $\R^3$, $X_{\theta}$ is defined
in (1.13).

We use that
$$\eqalignno{
&\mu^+(\cdot,\theta,\omega)=Id+D_{-\theta}(F(\cdot,\theta,\omega)
\mu^+(\cdot,\theta,\omega))\ \ {\rm on}\ \ \R^3,&(4.15)\cr
&S(\cdot,\theta,\omega)=Id+I_{\theta}(F(\cdot,\theta,\omega)
\mu^+(\cdot,\theta,\omega))\ \ {\rm on}\ \ X_{\theta},&(4.16)\cr}$$
where $\mu^+$, $F$, $S$  are defined in Section 1 (see (1.10), (1.11),
(1.12)),  $\theta\in\S^1_{\omega}$, $\omega\in\S^2$. In addition,
(4.15) is an integral equation for $\mu^+$, (4.16) is a formula for $S$.

\vskip 2 mm
{\bf Lemma 4.1.}
{\it Let} $u\in \hat C^{0,\sigma}(\R^3)$ {\it for some} $\sigma>3$. {\it Then}
$$\eqalign{
&I_{\theta}u\in \hat C^{0,\sigma}(X_{\theta}),\cr
&\|I_{\theta}u\|_{\hat C^{0,\sigma}(X_{\theta})}\le
2\pi\|u\|_{\hat C^{0,\sigma}(\R^3)},\ \ \theta\in\S^2,\cr}\eqno(4.17)$$
{\it where}
$$\eqalignno{
&\hat C^{0,\sigma}(X_{\theta})=\{g:\ \ \hat g\in C(X_{\theta}),\ \
\|g\|_{\hat C^{0,\sigma}(X_{\theta})}< +\infty\},&(4.18)\cr
&\hat g(p)=\bigl({1\over 2\pi}\bigr)^2\int\limits_{X_{\theta}}e^{ipx}g(x)dx,\
\ p\in X_{\theta},&(4.19)\cr}$$
$$\eqalign{
&\|g\|_{\hat C^{0,\sigma}(X_{\theta})}=
\|\hat g\|_{C^{0,\sigma}(X_{\theta})},\cr
&\|\hat g\|_{C^{0,\sigma}(X_{\theta})}=\sup_{p\in X_{\theta}}
(1+|p|)^{\sigma}|\hat g(p)|.\cr}\eqno(4.20)$$

Lemma 4.1 follows from the formula
$$(2\pi)^{-1}\hat g=\hat u\big|_{X_{\theta}}\ \ {\rm for}\ \ g=I_{\theta}u
\eqno(4.21)$$
and definitions (3.2)-(3.6), (4.18)-(4.20).

\vskip 2 mm
{\bf Lemma 4.2.}
{\it Let}
$$v(x,\theta,\omega)=\pmatrix{
\omega u(x)\omega \ &\ \omega u(x)\theta^{\perp}\cr
\theta^{\perp}u(x)\omega\ &\ \theta^{\perp}u(x)\theta^{\perp}\cr},\eqno(4.22)
$$
{\it where} $u$ {\it is a} ${\cal M}_{3,3}$- {\it valued function on}
$\R^3$, $x\in\R^3$, $\omega$, $\theta$, $\theta^{\perp}$ {\it are
vectors of} (1.7). {\it Then}
$$\eqalignno{
&\|v(\cdot,\theta,\omega)\|_{\hat L^{\infty,\sigma}(\R^3)}\le
c_2\|u\|_{\hat L^{\infty,\sigma}(\R^3)}\ \ {\it for}\ \ u\in
\hat L^{\infty,\sigma}(\R^3),\ \sigma\ge 0,&(4.23)\cr
&\|v(\cdot,\theta,\omega)\|_{\hat C^{\alpha,\sigma}(\R^3)}\le
c_2\|u\|_{\hat C^{\alpha,\sigma}(\R^3)}\ \ {\it for}\ \ u\in
\hat C^{\alpha,\sigma}(\R^3),\ \alpha\ge 0,\
\sigma\ge 0,&(4.24)\cr}$$
{\it where} $c_2$ {\it is some positive constant}.

Lemma 4.2 follows from definitions (4.22), (3.1)-(3.8).

\vskip 2 mm
{\bf Lemma 4.3.}
{\it Let} $u\in \hat C^{\alpha,\sigma}(\R^3)$ {\it for some}
$\alpha\in ]0,1[$, $\sigma>3$.
{\it Let} $v\in \hat C^{1+\alpha,\sigma}(\R^3)$.
{\it Then}
$$\eqalign{
&vD_{-\theta}u\in\hat C^{\alpha,\sigma}(\R^3),\cr
&\|vD_{-\theta}u\|_{\hat C^{\alpha,\sigma}(\R^3)}\le
c_3(\alpha,\sigma)\|v\|_{\hat C^{1+\alpha,\sigma}(\R^3)}
\|u\|_{\hat C^{\alpha,\sigma}(\R^3)},\ \ \theta\in\S^2,\cr}\eqno(4.25)$$
{\it for some positive}  $c_3=c_3(\alpha,\sigma)$.

In Lemma 4.3, $u$, $v$ are matrix-valued, in general, where the matrix
product is defined in the standard way.

\vskip 2 mm
{\it Proof of Lemma 4.3.}
We use that
$$\eqalign{
&D_{-\theta}u(x)=G_{\theta}^+* u(x)=\int\limits_{\R^3}
G_{\theta}^+(x-y)u(y)dy,\cr
&G_{\theta}^+(x)=\delta(\omega x)\delta(\theta^{\perp} x)h(\theta x),\cr}
\eqno(4.26)$$
where $\omega$, $\theta$, $\theta^{\perp}$ are related as in (1.7),
$\delta$ is the Dirac function,
$$\eqalign{
&h(s)=1\ \ {\rm for}\ \ s>0,\cr
&h(s)=0\ \ {\rm for}\ \ s\le 0.\cr}\eqno(4.27)$$
Further in this proof we assume for simplicity that
$$\theta=e_1,\ \ \theta^{\perp}=e_2,\ \ \omega=e_3,\eqno(4.28)$$
 where $e_1$, $e_2$, $e_3$ is the basis in $\R^3$.

We use that
$$\widehat{u_1* u_2}=(2\pi)^3\hat u_1\hat u_2,\eqno(4.29)$$
where $u_1$, $u_2$ are the same that in (4.4).

We use also that, under assumptions (4.28),
$$\eqalign{
&\hat G_{\theta}^+(p)=(2\pi)^{-3}\int\limits_{\R^3}
e^{ipx}\delta(x_3)\delta(x_2)h(x_1)dx=\cr
&(2\pi)^{-2}{1\over 2\pi}\int\limits_{\R}
e^{ip_1x_1}h(x_1)dx_1=(2\pi)^{-2}{-1\over 2\pi i}
{1\over {p_1+i0}}=(2\pi)^{-3}{i\over {p_1+i0}},\ \ p\in\R^3.
\cr}\eqno(4.30)
$$
Due to (4.26)-(4.30), we have that
$$\widehat{D_{\theta}u}(p)={i\hat u(p)\over {p_1+i0}},\ \ p\in\R^3.
\eqno(4.31)$$
Due to (4.4), (4.31), we have that
$$\widehat{vD_{\theta}u}(p)=\int\limits_{\R^3}\hat v(p-p^{\prime})
{i\hat u(p^{\prime})\over {p_1^{\prime}+i0}}dp^{\prime},\ \ p\in\R^3.
\eqno(4.32)$$
To prove Lemma 4.3, it is sufficient to prove that
$$\eqalignno{
&|\widehat{vD_{\theta}u}(p)|\le
{c_{3,1}(\alpha,\sigma)\|\hat v\|_{C^{\alpha,\sigma}(\R^3)}
\|\hat u\|_{C^{\alpha,\sigma}(\R^3)}\over (1+|p|)^{\sigma}},&(4.33a)\cr
&\big|{\pa \widehat{vD_{\theta}u}(p)\over \pa p_j}\big|\le
{c_{3,2}(\alpha,\sigma)\|\hat v\|_{C^{1+\alpha,\sigma}(\R^3)}
\|\hat u\|_{C^{\alpha,\sigma}(\R^3)}\over (1+|p|)^{\sigma}},&(4.33b)\cr}$$
for $p\in\R^3$, $j=1,2,3$ and some positive $c_{3,1}$, $c_{3,2}$.

Proceeding from (4.32) we have that
$$\eqalignno{
&\widehat{vD_{\theta}u}(p)=
\biggl(\int\limits_{|p_1^{\prime}|<1}+\int\limits_{|p_1^{\prime}|\ge 1}\biggr)
\hat v(p-p^{\prime})
{i\hat u(p^{\prime})\over {p_1^{\prime}+i0}}dp^{\prime}=A(p)+B(p),&(4.34)\cr
&A(p)=\pi
\int\limits_{p_1^{\prime}=0}\hat v(p-p^{\prime})\hat u(p^{\prime})dp^{\prime}
+p.v.\int\limits_{|p_1^{\prime}|<1}\hat v(p-p^{\prime})
{i\hat u(p^{\prime})\over p_1^{\prime}}dp^{\prime}=A_1(p)+A_2(p),&(4.35)\cr}$$
$$\eqalign{
&|A_1(p)|\le\pi\int\limits_{p_1^{\prime}=0}
{\|\hat v\|_{C^{0,\sigma}(\R^3)}\|\hat u\|_{C^{0,\sigma}(\R^3)}
\over  (1+|p-p^{\prime}|)^{\sigma}(1+|p^{\prime}|)^{\sigma}}dp^{\prime}\le\cr
&c_{3,1,1}(\sigma)
\|\hat v\|_{C^{0,\sigma}(\R^3)}\|\hat u\|_{C^{0,\sigma}(\R^3)}
(1+|p|)^{-\sigma},\cr}\eqno(4.36)$$
$$\eqalign{
&|A_2(p)|\le Const(\sigma)\int\limits_{|p_1^{\prime}|<1}\biggl(
\int\limits_{p_1^{\prime}=0}
{\|\hat v\|_{C^{\alpha,\sigma}(\R^3)}\|\hat u\|_{C^{\alpha,\sigma}(\R^3)}
\over  (1+|p-p^{\prime}|)^{\sigma}(1+|p^{\prime}|)^{\sigma}}dp^{\prime}
\biggr){|p_1^{\prime}|^{\alpha}\over |p_1^{\prime}|}dp_1^{\prime}\le\cr
&c_{3,1,2}(\alpha,\sigma)
\|\hat v\|_{C^{\alpha,\sigma}(\R^3)}\|\hat u\|_{C^{\alpha,\sigma}(\R^3)}
(1+|p|)^{-\sigma},\cr}\eqno(4.37)$$
$$\eqalign{
&|B(p)|\le\int\limits_{|p_1^{\prime}|\ge 1}
{\|\hat v\|_{C^{0,\sigma}(\R^3)}\|\hat u\|_{C^{0,\sigma}(\R^3)}
\over  (1+|p-p^{\prime}|)^{\sigma}(1+|p^{\prime}|)^{\sigma}}dp^{\prime}\le\cr
&c_1(\sigma)
\|\hat v\|_{C^{0,\sigma}(\R^3)}\|\hat u\|_{C^{0,\sigma}(\R^3)}
(1+|p|)^{-\sigma},\cr}\eqno(4.38)$$
where $p\in\R^3$, $c_{3,1,1}$, $c_{3,1,2}$ are some positive constants,
$c_1$ is the constant of (4.6).

Estimate (4.33a) follows from (4.34)-(4.38).
Estimate (4.33b) follows from (4.34)-(4.38) with $\hat v(p)$ replaced by
$\pa\hat v(p)/\pa p_j$.

Lemma 4.3 is proved.

We continue the proof of Proposition 3.2.

Using (4.15), the property that
$$F(\cdot,\theta,\omega)=\chi F(\cdot,\theta,\omega),\eqno(4.39)$$
where $\chi$ is the function of (3.16), (3.17),
 and Lemmas 3.1, 4.3 we obtain that
$$\eqalign{
&\mu^+(\cdot,\theta,\omega)=Id+
\sum_{j=1}^{+\infty}D_{-\theta}w_j(\cdot,\theta,\omega),\cr
&w_j(\cdot,\theta,\omega)=\underbrace{F(\cdot,\theta,\omega)D_{-\theta}\ldots
F(\cdot,\theta,\omega)D_{-\theta}}_{j-1}F(\cdot,\theta,\omega),\cr}\eqno(4.40)
$$
$$F(\cdot,\theta,\omega)\mu^+(\cdot,\theta,\omega)=F(\cdot,\theta,\omega)+
\sum_{j=1}^{+\infty}w_{j+1}(\cdot,\theta,\omega),\eqno(4.41)$$
where
$$\eqalign{
&\|w_j(\cdot,\theta,\omega)\|_{\hat C^{\alpha,\sigma}(\R^3)}\le
\bigl(c_3(\alpha,\sigma)
\|F(\cdot,\theta,\omega)\|_{\hat C^{1+\alpha,\sigma}(\R^3)}\bigr)^{j-1}
\|F(\cdot,\theta,\omega)\|_{\hat C^{\alpha,\sigma}(\R^3)}\le\cr
&q_1(q_2)^{j-1}\ep^j,\cr}\eqno(4.42)$$
where
$$\eqalign{
&q_1=\lambda_1(\alpha,\sigma)c_2\|\chi\|_{\hat C^{\alpha,\sigma}(\R^3)},\cr
&q_2=\lambda_1(1+\alpha,\sigma)c_2c_3(\alpha,\sigma)
\|\chi\|_{\hat C^{1+\alpha,\sigma}(\R^3)},\cr}\eqno(4.43)$$
and $\lambda_1$, $c_2$, $c_3$, $\ep$ are the numbers of Lemmas 3.1, 4.2, 4.3
and Theorem 3.1.

Due to (4.41)-(4.43), we have  that
$$\eqalign{
&\|F(\cdot,\theta,\omega)\mu^+(\cdot,\theta,\omega)-
F(\cdot,\theta,\omega)\|_{\hat C^{\alpha,\sigma}(\R^3)}\le\cr
&q_1\ep\sum_{j=1}^{+\infty}(q_2\ep)^j={q_1q_2\ep^2\over {1-q_2\ep}},\cr}
\eqno(4.44)$$
under the condition that $q_2\ep<1$.

Due to (4.16), (4.44) and Lemma 4.1, we have that
$$\|S(\cdot,\theta,\omega)-Id-
I_{\theta}F(\cdot,\theta,\omega)\|_{\hat C^{0,\sigma}(X_{\theta})}\le
{2\pi q_1q_2\ep^2\over {1-q_2\ep}} \eqno(4.45)$$
and, in particular,
$$\|S_{11}(\cdot,\theta,\omega)-1-
I_{\theta}\omega f\omega\|_{\hat C^{0,\sigma}(X_{\theta})}\le
{2\pi q_1q_2\ep^2\over {1-q_2\ep}}. \eqno(4.46)$$
To obtain (3.26) we use (4.46), (4.43) and the following estimate
$$\|\chi\|_{\hat C^{\alpha,\sigma}(\R^3)}\le c_4(\alpha,\sigma)
\|\chi\|_{\hat C^{1+\alpha,\sigma}(\R^3)},\eqno(4.47)$$
where $c_4$ is an appropriate positive constant. In addition, it is assumed
that $\ep_0$ of

\noindent
Theorem 3.1 is so small that
$$q_2\ep_0\le 1/2.\eqno(4.48)$$
As a result, proceeding from (4.46) we obtain (3.24), (3.26).

Finally, (3.27) follows from (3.24)-(3.26) and Proposition 3.1.

Proposition 3.2 is proved.

\vskip 2 mm
{\bf 5. Proof of Proposition 3.3}

We will use that in the construction of $f^{n+1}$ from $f^n$ and $S_{11}$
the steps given by (2.13), (2.14) can be rewritten as
$$f^{n+1}=\chi J^{-1}_{\Lambda}(S_{11}-1-T_{11}^n),\eqno(5.1)$$
where  $T_{11}^n$ is the element of $T^n=(T_{ij}^n)$, where
$$T^n(x,\theta,\omega)=\lim\limits_{s\to +\infty}U^n(x+s\theta,\theta,\omega),
\ \  (x,\theta,\omega)\in\Lambda,\eqno(5.2)$$
$$\eqalign{
&\theta\pa_xU^n(x,\theta,\omega)=F^n(x,\theta,\omega)
(\mu^{n+}(x,\theta,\omega)-Id),\cr
&\lim\limits_{s\to -\infty}U^n(x+s\theta,\theta,\omega)=0,
\ \  (x,\theta,\omega)\in {\cal V},\cr}\eqno(5.3)$$
where $F^n$ and $\mu^{n+}$ are defined as $F$ and $\mu^+$ of (1.10), (1.11)
with $f^n$ in place of $f$, $\cal V$ is defined by (2.11).

Indeed, (5.2), (5.3) and the definition of $\mu^{n+}$ imply that
$$\eqalignno{
&U^n(x,\theta,\omega)=\mu^{n+}(x,\theta,\omega)-Id-D_{-\theta}
F^n(x,\theta,\omega),\ \ (x,\theta,\omega)\in {\cal V},&(5.4)\cr
&T^n_{11}(x,\theta,\omega)=S^n_{11}(x,\theta,\omega)-1-I_{\theta}\omega
f^n(x)\omega,\ \ (x,\theta,\omega)\in \Lambda,&(5.5)\cr}$$
where $I_{\theta}$ and $D_{\theta}$ are defined by (4.13), (4.14).
Using (5.5), (2.5)
one can see that (5.1) is equivalent to
$$f^{n+1}=\chi J^{-1}_{\Lambda}(S_{11}-S^n_{11}+J f^n).\eqno(5.6)$$
Using that  $J^{-1}_{\Lambda} J f^n=f^n$ (this is completely similar to
(3.23)), one can see that $f^{n+1}$ of (5.1), (5.6) coincides with
$f^{n+1}$ of (2.14).

Thus, it is sufficient to prove Proposition 3.3, where  (2.14) is
written as (5.1).

In this proof, in addition to $T^n$ and $U^n$, we consider also $T$ and $U$
(defined in a completely similar way with $T^n$, $U^n$):
$$T(x,\theta,\omega)=\lim\limits_{s\to +\infty}U(x+s\theta,\theta,\omega),
\ \  (x,\theta,\omega)\in\Lambda,\eqno(5.7)$$
$$\eqalign{
&\theta\pa_xU(x,\theta,\omega)=F(x,\theta,\omega)
(\mu^+(x,\theta,\omega)-Id),\cr
&\lim\limits_{s\to -\infty}U(x+s\theta,\theta,\omega)=0,
\ \  (x,\theta,\omega)\in {\cal V},\cr}\eqno(5.8)$$
In addition, (in a completely similar way with (5.4), (5.5)) we have that
$$\eqalignno{
&U(x,\theta,\omega)=\mu^+(x,\theta,\omega)-Id-D_{-\theta}
F(x,\theta,\omega),\ \ (x,\theta,\omega)\in {\cal V},&(5.9)\cr
&T_{11}(x,\theta,\omega)=S_{11}(x,\theta,\omega)-1-I_{\theta}\omega
f^n(x)\omega,\ \ (x,\theta,\omega)\in \Lambda.&(5.10)\cr}$$

One can see that
$$\eqalign{
&f^{n+1}\buildrel (5.1) \over = \chi J^{-1}_{\Lambda}(S_{11}-1-T_{11})+
\chi J^{-1}_{\Lambda}(T_{11}-T^n_{11})\buildrel (2.5), (5.10) \over =\cr
&\chi J^{-1}_{\Lambda}(J f)+
\chi J^{-1}_{\Lambda}(T_{11}-T^n_{11})\buildrel (3.23) \over =
\chi(f+\delta^{n+1}f),\cr}\eqno(5.11)$$
where
$$\delta^{n+1}f=J^{-1}_{\Lambda}(T_{11}-T^n_{11}).\eqno(5.12)$$
Thus, to complete the proof of Proposition 3.3 it is sufficient to obtain
an appropriate estimate on $T_{11}-T^n_{11}$ (estimate (5.31) given below).

Let
$$\eqalignno{
&\delta^nT_{11}=T^n_{11}-T_{11},&(5.13)\cr
&\delta^nF=F^n-F,&(5.14)\cr
&\delta^n\mu^+=\mu^{n+}-\mu^+.&(5.15)\cr}$$
Using the definitions of $T^n$ and $T$ one can see that
$\delta^nT_{11}$ is the element of
$\delta^nT=(\delta^nT_{ij})$, where
$$\delta^nT=I_{\theta}(F^n(\cdot,\theta,\omega)(\mu^{n+}(\cdot,\theta,\omega)
-Id)-
F(\cdot,\theta,\omega)(\mu^+(\cdot,\theta,\omega)-Id)),\eqno(5.16)$$
where $I_{\theta}$ is defined by (4.13). In addition, using (5.14), (5.15)
formula (5.16) can be rewritten as
$$\delta^nT=I_{\theta}(\delta^nF(\cdot,\theta,\omega)
(\mu^+(\cdot,\theta,\omega)-Id)+
F^n(\cdot,\theta,\omega)\delta^n\mu^+(\cdot,\theta,\omega)).\eqno(5.17)$$
The following estimates hold:
$$\eqalign{
&\|\delta^nF(\cdot,\theta,\omega)
(\mu^+(\cdot,\theta,\omega)-Id)\|_{\hat C^{\alpha,\sigma}(\R^3)}\le\cr
&\lambda_1(\alpha,\sigma)c_2
\bigl({q_1q_3\ep\over {1-q_2\ep}}\bigr)
\|\delta^nf\|_{\hat L^{\infty,\sigma}(\R^3)},\cr}\eqno(5.18)$$
$$\eqalign{
&\|F^n(\cdot,\theta,\omega)\delta^n
\mu^+(\cdot,\theta,\omega)\|_{\hat C^{\alpha,\sigma}(\R^3)}\le\cr
&{q_1q_2\ep\|\delta^nf\|_{\hat L^{\infty,\sigma}(\R^3)}
\over (1-q_2\ep)^2},\cr}\eqno(5.19)$$
under the condition that $q_2\ep<1$, where $q_1$, $q_2$ are given by (4.43),
$$q_3=c_3(\alpha,\sigma)\|\chi\|_{\hat C^{1+\alpha,\sigma}(\R^3)},\eqno(5.20)
$$
$\ep$ is the number of (3.29), $\lambda_1$, $c_2$, $c_3$ are the numbers of
Lemmas 3.1, 4.2, 4.3, $\theta\in\S^1_{\omega}$, $\omega\in\S^2$.

\vskip 2 mm
{\it Proof of (5.18).}
Using (3.28), (4.39) and the definiions of $F$, $F^n$, $\delta^nF$ we have
that
$$\delta^nF(x,\theta,\omega)=\chi(x)\pmatrix{
\omega\delta^nf(x)\omega\ &\ \omega\delta^nf(x)\theta^{\perp}\cr
\theta^{\perp}\delta^nf(x)\omega\ &\ \theta^{\perp}\delta^nf(x)
\theta^{\perp}\cr},\eqno(5.21)$$
$x\in\R^3$, $\theta\in\S^1_{\omega}$, $\omega\in\S^2$.

Using (4.40), (4.42) and Lemma 4.3 we obtain that
$$\|\chi(\mu^+(\cdot,\theta,\omega)-Id)\|_{\hat C^{\alpha,\sigma}(\R^3)}\le
q_1q_3\ep\sum_{j=1}^{+\infty}(q_2\ep)^{j-1}={q_1q_3\ep\over {1-q_2\ep}}
\eqno(5.22)$$
under the condition that $q_2\ep<1$.

Estimate (5.18) follows from (5.21), (5.22) and Lemmas 3.1 and 4.2.

\vskip 2 mm
{\it Proof of (5.19).}
Using (4.40) for $\mu^+$ and for $\mu^{n+}$ we have that
$$\eqalign{
&F^n(\cdot,\theta,\omega)\delta^n\mu^+(\cdot,\theta,\omega)=
F^n(\cdot,\theta,\omega)\sum_{j=1}^{+\infty}D_{-\theta}
\delta^nw_j(\cdot,\theta,\omega),\cr
&\delta^nw_j(\cdot,\theta,\omega)=w_j^n(\cdot,\theta,\omega)-
w_j(\cdot,\theta,\omega),\cr}\eqno(5.23)$$
where $w_j^n$ is defined as $w_j$ of (4.40), but with $F^n$ in place of
$F$. In addition, one can see that
$$\eqalign{
&\delta^nw_1(\cdot,\theta,\omega)=\delta^nF(\cdot,\theta,\omega),\cr
&\delta^nw_{j+1}(\cdot,\theta,\omega)=
\delta^nF(\cdot,\theta,\omega)D_{-\theta}w^n_j(\cdot,\theta,\omega)+
F(\cdot,\theta,\omega)D_{-\theta}\delta^nw_j(\cdot,\theta,\omega),\ j\in\N.\cr
}\eqno(5.24)$$
In addition, using (5.21), (5.24), Lemma 4.3, estimate (4.42) for
$w_j^n$ and Lemma 3.1 we have that
$$\eqalign{
&\|\delta^nw_{j+1}(\cdot,\theta,\omega)\|_{\hat C^{\alpha,\sigma}(\R^3)}\le\cr
&c_3(\alpha,\sigma)
\|\delta^nF(\cdot,\theta,\omega)\|_{\hat C^{1+\alpha,\sigma}(\R^3)}
q_1q_2^{j-1}\ep^j+\cr
&c_3(\alpha,\sigma)
\|F(\cdot,\theta,\omega)\|_{\hat C^{1+\alpha,\sigma}(\R^3)}
\|\delta^nw_j(\cdot,\theta,\omega)\|_{\hat C^{\alpha,\sigma}(\R^3)}\le\cr
&q_1q_2^j\ep^j\|\delta^nf\|_{\hat L^{\infty,\sigma}(\R^3)}+
q_2\ep\|\delta^nw_j(\cdot,\theta,\omega)\|_{\hat C^{\alpha,\sigma}(\R^3)},\cr
&\|\delta^nw_1(\cdot,\theta,\omega)\|_{\hat C^{\alpha,\sigma}(\R^3)}\le
q_1\|\delta^nf\|_{\hat L^{\infty,\sigma}(\R^3)}.\cr}\eqno(5.25)$$
Using (5.25) we obtain that
$$\eqalign{
&\|\delta^nw_j(\cdot,\theta,\omega)\|_{\hat C^{\alpha,\sigma}(\R^3)}\le
q_1\|\delta^nf\|_{\hat L^{\infty,\sigma}(\R^3)}Z_j,\cr
&Z_{j+1}=(q_2\ep)^j+q_2\ep Z_j,\ \ Z_1=1.\cr}\eqno(5.26)$$

In addition, one can see that
$$Z_j=j(q_2\ep)^{j-1},\ \ j\in\N.\eqno(5.27)$$

Estimate (5.19) follows from (5.23), (5.26), (5.27), Lemmas 3.1 and 4.3 and
the formula
$$\sum_{j=1}^{+\infty}jr^{j-1}={1\over (1-r)^2},\ \ 0<r<1.\eqno(5.28)$$
Using (5.12), (5.18), (5.19) and Lemma 4.1 we obtain that
$$\eqalign{
&\|\delta^nT(\cdot,\theta,\omega)\|_{\hat C^{0,\sigma}(X_{\theta})}\le\cr
&\bigl({\lambda_1c_2q_1q_3\over {1-q_2\ep}}+
{q_1q_2\over (1-q_2\ep)^2}\bigr)\ep
\|\delta^nf\|_{\hat L^{\infty,\sigma}(\R^3)},\ \ \theta\in\S^1_{\omega},\ \
\omega\in\S^2.\cr}\eqno(5.29)$$
Proceeding from (5.29) and assuming that $\ep_1$ of (3.29) is so small that
$$q_2\ep_1\le 1/2 \eqno(5.30)$$
we obtain that
$$\|\delta^nT\|_{\hat L^{\infty,\sigma}(\Lambda)}\le
(2\lambda_1c_2q_1q_3+4q_1q_2)\ep
\|\delta^nf\|_{\hat L^{\infty,\sigma}(\R^3)},\eqno(5.31)$$
where $q_1$, $q_2$, $q_3$, $\lambda_1$, $c_2$ are the same that in (5.18),
(5.19).

Finally, (3.31) follows from (5.12), (5.31), (4.47)  and Proposition 3.1.

Proposition 3.3 is proved.

\vskip 2 mm
{\bf 6. Final remarks}

\vskip 2 mm
{\bf Remark 6.1.}
In a subsequent paper we plan to generalize the iterative approach of the
present work to the case of the polarization tomography with limited phase
measurements, see, for example, [HL], [Sh3] for more information on this
problem.
Actually, in the framework of the model described by (1.1) the polarization
tomography with limited phase measurements is reduced to the inverse
problem for (1.10) with $f-(1/2) tr\,(\pi_{\theta}f\pi_{\theta}) Id$ in
place of $f$. In this inverse problem we do not plan to restrict the
"scattering" matrix $S$ to its element $S_{11}$ only (in contrast with
results of the present work).

\vskip 2 mm
{\bf Remark 6.2.}
It remains unclear whether a version of the Riemann-Hilbert problem method
of [MZ], [N] can be used for solving Problem 1.1, instead of the
iterative approach of the present work. The reason is that the dependence
of $F(x,\theta,\omega)$ on the spectral parameter $\theta$ (and, more
precisely, the quadratic dependence on $\theta$ of the element
$F_{22}(x,\theta,\omega)=\theta^{\perp}f(x)\theta^{\perp}$) in (1.10) is
not appropriate, in general, for direct applications of
the Riemann-Hilbert problem method of [MZ], [N].

\vskip 2 mm
{\bf Remark 6.3.}
On the other hand (with respect to Remark 1.2), if $f$ is skew-symmetric,
$f_{ij}=-f_{ji}$, then $F_{22}\equiv 0$ and the dependence of
$F(x,\theta,\omega)$ on $\theta$ is appropriate for direct applications of
aforementioned  Riemann-Hilbert problem method to the inverse problem for
equations (1.1), (1.10). We remind that some results on the polarization
tomography with skew-symmetric $f$, including examples of transparent $f$,
were given in [NS]. However, the Riemann-Hilbert problem method was not
yet used in these studies.

\vskip 2 mm
{\bf References}
\item{[  A]} H.K. Aben, Integrated Photoelasticity (New York: McGraw-Hill),
1979.
\item{[ DP]} M. Dunajski, P. Plansangkate, Topology and energy of time
dependent unitons, Proc. R. Soc. A {\bf 463} (2007), 945-959.
\item{[  E]} G. Eskin, On non-abelian Radon transform, Russ.J.Math.Phys.
{\bf 11} (2004), 391-408.
\item{[ FU]} D. Finch, G. Uhlmann, The X-ray transform for a non-abelian
connection in two dimensions, Inverse Problems {\bf 17} (2001), 695-701.
\item{[ HL]} H.B. Hammer and W.R. Lionheart, Reconstruction of spatially
inhomogeneous dielectric tensors through optical tomography, J. Opt.Soc.Am.
{\bf 22} (2005), 250-255.
\item{[ KO]} Yu.A. Kravtsov and Yu.I. Orlov, Geometric Optics of Inhomogeneous
Media (Moscow: Nauka) (in Russian), 1980.
\item{[  M]} L.J. Mason, Global anti-self dual Yang-Mills fields in split
signature and their scattering, J.Reine Angew.Math. {\bf 597} (2006),
105-133.
\item{[ MZ]} S.V. Manakov and V.E. Zakharov, Three-dimensional model of
relativistic-invariant field theory, integrable by inverse scattering
transform, Lett.Math.Phys. {\bf 5} (1981), 247-253.
\item{[  N]} R.G. Novikov, On determination of a gauge field on $\R^3$ from
its non-abelian Radon transform along oriented straight lines, J.Inst.Math.
Jussieu {\bf 1} (2002), 559-629.
\item{[ NS]} R.G. Novikov and V.A. Sharafutdinov, On the problem of
polarization tomography: I, Inverse Problems {\bf 23} (2007), 1229-1257.
\item{[  P]} G.P. Paternain, Transparent connections over negatively curved
surfaces,

\item{    } arXiv:0809.4360v3.
\item{[Sh1]} V.A. Sharafutdinov, Integral Geometry of Tensor Fields
(Utrecht: VSP), 1994.
\item{[Sh2]} V.A. Sharafutdinov, On an inverse problem of determining a
connection on a vector bundle, J.Inverse and Ill-Posed Problems {\bf 8}
(2000), 51-88.
\item{[Sh3]}  V.A. Sharafutdinov, The problem of polarization tomography: II,
Inverse Problems {\bf 24} (2008), 035010 (21pp).
\item{[  V]}  L.B. Vertgeim, Integral geometry with a matrix weight, and a
nonlinear problem of recovering matrices, Sov.Math.Dokl. {\bf 44} (1992),
132-135.

\end